\documentclass[prl,amsmath,amssymb,epsfig,twocolumn, showpacs]{revtex4}

\usepackage{graphicx}
\usepackage{dcolumn}
\usepackage{bm}

\begin{document}

\title{Translationally invariant discrete kinks from one-dimensional
maps}
\author{I.V. Barashenkov}
\author{O.F. Oxtoby}
\affiliation{Department of Mathematics, University of
Cape Town, Rondebosch 7701,
South Africa}
\author{Dmitry E. Pelinovsky}
\affiliation{Department of Mathematics, McMaster
University, Hamilton, Ontario, Canada, L8S 4K1}

\date{\today}
 
\begin{abstract}
For most discretisations of the $\phi^4$ theory, the
stationary kink can only be centered either on a lattice site or
midway between two adjacent sites. We search for 
exceptional discretisations which allow stationary kinks to be
centered anywhere between the sites. We show that this translational
invariance of the kink implies the existence of an underlying
one-dimensional map $\phi_{n+1}=F(\phi_n)$. A simple algorithm based 
on this observation generates three different  families of
exceptional discretisations.

\end{abstract}

\pacs{05.45.Yv, 63.20.Pw}

\maketitle

Since the early 1960s, the  $\phi^4$-equation,
\begin{equation}
\phi_{tt} = \phi_{xx} + \frac{1}{2} \phi ( 1 -
\phi^2),
\label{continuous-phi4} 
\end{equation} 
has been 
one of the workhorses of statistical mechanics \cite{Bishop_Schneider} and 
quantum field theory
\cite{Rajaraman}.  
Its kink solution,
\begin{equation}
\label{travelling} 
\phi(x,t) = \tanh \frac{x - ct - x^{(0)}}{2 \sqrt{1
- c^2}},
\end{equation} 
together with the sine-Gordon kink, 
are the simplest examples of topological solitons.
More recently, interest has shifted towards the 
{\it discrete\/}  $\phi^4$-theories
\cite{Sanchez,phi4kinks}, 
\begin{equation}
\label{phi-4} 
\ddot{\phi}_n = \frac{\phi_{n+1} - 2 \phi_n + \phi_{n-1}}{h^2} +
f(\phi_{n-1}, \phi_n, \phi_{n+1}),
\end{equation}
and their solutions.
Here  $h$ is 
the lattice spacing: $\phi_n(t)=\phi(x_n,t)$, with $x_n=hn$, and
 the function $f$ is chosen
to reproduce the nonlinearity (\ref{continuous-phi4}) 
in the continuum limit:
\begin{equation}
 f(\phi,\phi,\phi) = \frac{1}{2} \phi (1
- \phi^2).
\label{flim}
\end{equation}

For a variety of discrete nonlinearities $f$, Eq.(\ref{phi-4}) admits 
stationary kink solutions \cite{Slovenians,Yip}. 
The discrete  $\phi^4$ kinks have been used to describe incommensurate 
systems and narrow 
domain walls in ferroelectrics and ferromagnets, 
 topological excitations in 
biological macromolecules and hydrogen-bonded chains, 
and polymerization mismatches in polymers
\cite{Sanchez,Applications}.
Physically, one of the 
most significant properties of domain walls
and topological defects is their mobility
\cite{Yip};
however whether the discrete  equations (\ref{phi-4})
  can admit 
travelling kinks remains an open
question
\cite{phi4kinks,moving_kinks,Speight1,Speight2,Speight3,Panos}. 
The continuous $\phi^4$-equation (\ref{continuous-phi4}) 
is Lorentz-invariant,
and so the existence of the travelling kink (\ref{travelling}) 
is an immediate consequence of the existence of the 
 stationary
soliton, Eq.(\ref{travelling}) with $c=0$. 
The discretisation breaks 
 the Lorentz  invariance and the 
existence of  travelling discrete kinks becomes a nontrivial matter.

In fact, the discretisation 
 even breaks  the translation
invariance of Eq.(\ref{continuous-phi4}). As a result, the
stationary kink can be centered only at a countable number
of points --- usually  
 on a site and midway between  two adjacent sites
 \cite{Slovenians,Yip}. 
This breaking of the translation invariance is 
connected with the presence of the Peierls-Nabarro barrier, 
an additional
periodic potential induced by  discreteness. 

Miraculously, there are several {\it exceptional\/}
 discretisations which, while breaking the translation
 invariance of the equation, allow the existence of 
 translationally invariant kinks; that is, 
 kinks centred at an arbitrary point between the sites.
One such  discretisation was discovered
 by Speight and Ward using a Bogomolny-type energy-minimality
  argument \cite{Speight1,Speight2}:
\begin{eqnarray}
 f 
=\frac{ 2\phi_n +
\phi_{n+1}}{12} \left( 1 - \frac{\phi_n^2 + \phi_n \phi_{n+1} 
+ \phi_{n+1}^2}{3}
\right) \nonumber \\
  +  \frac{2 \phi_n + \phi_{n-1}}{12}
\left( 1 - \frac{\phi_n^2 + \phi_n \phi_{n-1} + \phi_{n-1}^2}{3} \right).
\label{nonlinearity-Speight}
\end{eqnarray}
Another one derives from the Ablowitz-Ladik 
integrable discretisation of the nonlinear Schr\"odinger 
equation; it was reobtained  by Bender and Tovbis \cite{BT97}
from the requirement of suppression
of the kinks'
resonant radiation:
\begin{equation}
\label{nonlinearity-Tovbis} 
f = 
\frac{1}{4}
\left( \phi_{n+1} + \phi_{n-1} \right) \left( 1 - \phi_n^2 \right).
\end{equation}
Finally, the nonlinearity 
\begin{eqnarray}
\label{nonlinearity-Panos} f = \frac{
\phi_{n+1} + \phi_{n-1}}{4} -\frac{ ( \phi_{n+1}^2 +
\phi_{n-1}^2 ) (
\phi_{n+1} + \phi_{n-1} )}{8}
\end{eqnarray}
was identified by Kevrekidis \cite{Panos}, who demonstrated the existence
of a two-point invariant  associated
with the stationary equation 
\begin{equation}
\label{stat} 
\frac{\phi_{n+1} - 2 \phi_n + \phi_{n-1}}{h^2} +
f(\phi_{n-1}, \phi_n, \phi_{n+1})=0,
\end{equation}
 with $f$ as in (\ref{nonlinearity-Tovbis}) and
 (\ref{nonlinearity-Panos}).

Although the translation invariance
of a stationary kink does not automatically
guarantee the existence of a travelling soliton, it is natural
to expect it to be a prerequisite for kink mobility.
For example, in the variational description
of the slowly moving
kink, the  solution
is sought as a stationary kink 
with a free continuous parameter defining its position on
the line 
 \cite{Speight1}. Also,  the Stokes constants measuring the
 intensity
 of  resonant radiation from  the translationally
 invariant kinks were found to be 
  at least an order of magnitude smaller 
   than the corresponding constants in  models with noninvariant kinks \cite{ODI}.
 With an eye to a future attack on
 travelling kinks, 
 it would be useful to
 identify all
  discretisations of the $\phi^4$ theory
  supporting translationally invariant stationary kinks.
The purpose of this note is to provide a general
recipe for the generation of such exceptional discretisations
$f(\phi_{n-1}, \phi_n, \phi_{n+1})$.

We start with a simple observation which, however, holds the key 
to our construction. Assume we have a nonlinearity $f$
which supports a stationary discrete kink, i.e. a monotonically
growing sequence $\phi_n$: $-1<\phi_n<\phi_{n+1} <1$ for
$-\infty < n < \infty$. 
Furthermore, assume there exists 
 a continuous monotonically
growing function $g(x)$, defined for all real $x$, 
such that $\phi_n=g(n)$. (The function 
$g$ can also depend on $h$ parametrically but
we omit this dependence for simplicity of
notation.) This function
generates a family of kinks centered 
at an arbitrary point $x^{(0)}$ on the
line: $\phi_n=g(n-x^{(0)})$. 
It is important to emphasise that such a continuous function can
exist only
for exceptional discretisations.
For generic discretisations, the function 
$g(x)$ can be defined only on integers.

The existence of the function $g(x)$ defined on the entire real
line --- or,
equivalently, the translation invariance of the kink ---
 implies that the stationary equation (\ref{stat}) derives from
  a two-point map.  Indeed, since $g(x)$ 
is monotonic, we can write
$n=g^{-1}(\phi_n)$.
Now since $\phi_{n+1}=g(n+1)$, we have 
$\phi_{n+1}=g(g^{-1}(\phi_n) +1) \equiv F(\phi_n)$,
which is a well-defined one-dimensional map.

This observation suggests the following strategy for the construction 
of exceptional discretisations. Assume we have a 1D map
which we will  write in the form
\begin{equation}
\phi_{n+1}-\phi_{n}=hH(\phi_{n+1}, \phi_{n}).
\label{m5}
\end{equation}
 Let $H$  satisfy the following continuity condition:
\begin{equation}
H(\phi,\phi)= \frac12 (1-\phi^2).
\label{continuity}
\end{equation}
This condition is necessary to make sure that the map (\ref{m5})
becomes 
\begin{equation}
\phi_x=\frac12(1-\phi^2)
\label{Bogomolny}
\end{equation} 
in the continuum limit. The stationary ($c=0$) kink
solution (\ref{travelling}) of Eq.(\ref{continuous-phi4}) is,
simultaneously, a solution of the first-order equation
(\ref{Bogomolny}).
Imposing (\ref{continuity}) we ensure that the discrete kink of 
(\ref{m5}) will have the correct continuum limit.
Next, Eq.(\ref{continuity}) implies that
the map  (\ref{m5})
has just one pair of fixed points, 
$\phi_*= \mp 1$. 
For small $h$, $\phi_{n+1}$ remains close to $\phi_n$ 
and hence, $H(\phi_{n+1},\phi_n)$ remains close to (\ref{continuity})
which is positive for $|\phi|<1$. Consequently, no
matter what $|\phi_0|<1$ we start with, the sequence $\phi_n$ is 
monotonically growing --- at least until $|\phi_n|$
is not very close to 1.
To ensure that it remains such near the fixed
points, we assume that $\phi_*=-1$ is a source
 and $\phi_*=1$ a sink.
(That is, small 
perturbations $\delta \phi_n=\phi_n- \phi_*$ satisfy 
$\delta \phi_{n+1}=\lambda \delta \phi_n$ with $\lambda >1$
near $\phi_*=-1$ and $0<\lambda<1$ near $\phi_*=1$.)
Then, 
for any $h$ smaller than some $\overline{h}$ and any $\phi_0$ between
$-1$ and $1$, there is a number $N$ such that $|\phi_N-\phi_*|$ is
so small that all $\phi_n$ with $n>N$ are 
entrapped by the ``linear  neighbourhood" of
 $\phi_*=1$ and those with $n<-N$ are all in a neighbourhood of
  $\phi_* =-1$.
This means that each $\phi_0$ with $|\phi_0|<1$ 
defines  a monotonic kink solution 
and so for any sufficiently small $h$ we have a one-parameter 
family of stationary kinks.
Speight \cite{Speight3} gives a less intuitively
appealing but more rigorous proof of this
fact;
he also shows that our assumption on the character
of the fixed points can be relaxed.

Next, squaring both sides of (\ref{m5}) and
subtracting the square of 
its back-iterated copy,
\begin{equation}
\phi_{n}-\phi_{n-1}=hH(\phi_{n}, \phi_{n-1}),
\label{m2}
\end{equation}
produces an exceptional  stationary  Klein-Gordon equation
\begin{equation}
\frac{\phi_{n+1}-2\phi_{n}+\phi_{n-1}}{h^2}
=\frac{H^2(\phi_{n+1},\phi_n)
-H^2(\phi_n,\phi_{n-1})}{\phi_{n+1}-\phi_{n-1}}.
\label{m3}
\end{equation} 
If $H$ is symmetric:
$H(\phi_{n}, \phi_{n-1})=H(\phi_{n-1}, \phi_{n})$,
the numerator vanishes exactly where the denominator equals zero, so
the discretisation (\ref{m3}) is nonsingular.

If we want to have polynomial discretisations
of the $\phi^4$ theory, the function $H^2$ 
has to be a quartic polynomial. This leads to two possibilities,
one where $H$ is the square root of a polynomial, and the other where
$H$ is a polynomial itself.
These can be written jointly as
\begin{equation}
\left( \phi_{n+1}-\phi_n \right)^m = h^m P_{2m}(\phi_{n+1}, \phi_n),
 \label{s5}
\end{equation}
where $m=1$ or $2$, and $P_{2m}(u,v)$ is a polynomial of degree $2m$
that satisfies the symmetry and continuity conditions
\begin{eqnarray}
P_{2m}(u,v) & = & P_{2m}(v,u),
 \label{as1} \\ 
 P_{2m}(\phi,\phi) & = & 2^{-m} (1-\phi^2)^m. \label{as2}
 \end{eqnarray}
 The condition (\ref{as2}) is a consequence
 of Eq.(\ref{continuity}).

Before we proceed to the classification of the resulting
models, it is pertinent to note that the linear part
of the function
$f$ in (\ref{phi-4}) 
can always be fixed to $\frac12 \phi_n$ without
loss of generality.  Indeed, the most general function satisfying
\eqref{flim} is $f = a\phi_n +
\frac12\left(\frac12-a\right)\left(\phi_{n+1}
+\phi_{n-1}\right)+\text{cubic terms}$. 
Since $h^2$ in \eqref{stat} is a free parameter, 
we can always make a
replacement $h \to \tilde{h}$ such that 
$a-2/h^2 =\tilde{a}- 2/\tilde{h}^2$.  In
particular, we can set $\tilde{a} = \frac12$ which gives
\begin{equation}
f(\phi_{n-1}, \phi_n, \phi_{n+1}) = \frac12 \phi_n - Q(\phi_{n-1}, 
\phi_n, \phi_{n+1}), 
\label{canonical-form}
\end{equation}
where $Q$ is a homogeneous polynomial of degree 3.

Let, now, $m=2$ in Eq.(\ref{s5}).
Provided $P_4$ satisfies conditions (\ref{as1}) and (\ref{as2}),
the numerator $P_4(\phi_{n+1},\phi_n)-P_4(\phi_n,\phi_{n-1})$
of the fraction in the right-hand side of Eq.(\ref{m3})
divides $(\phi_{n+1}-\phi_{n-1})$ and so   Eq.(\ref{m3}) will be
 of the form
(\ref{stat}) with some cubic function $f$. The most general choice 
for such a polynomial is 
\begin{multline}
P_4(u,v)=\frac14 - \mu (u-v)^2 - \frac12 uv 
 \\ 
+ \frac{1}{20} \left[
\alpha (u^4 + v^4) + \beta u v (u^2 + v^2)
+ \gamma u^2 v^2 \right], 
\label{s8}
\end{multline}
where $\alpha$, $\beta$, $\gamma$ satisfy
$2\alpha + 2\beta + \gamma = 5$
and $\mu$
is arbitrary. Picking the positive value of 
$\sqrt{P_4}$ and assuming that $h$ is 
sufficiently small,
one can 
check that the fixed 
points $\phi_*=\mp 1$ of the map (\ref{s5}) are a source and 
a sink, for any $\mu, \alpha$ and $\beta$.  
Consequently, the resulting cubic polynomial,
\begin{multline}
 Q = \frac{1}{20} \left[ \alpha(\phi_{n+1}+\phi_{n-1})
(\phi_{n+1}^2+ \phi_{n-1}^2) \right. 
+ \gamma \phi_n^2(\phi_{n+1}  \\  
+ \phi_{n-1})  
\left.
+ \beta \phi_n(\phi_{n+1}^2 +\phi_n^2 
+\phi_{n-1}^2 +\phi_{n+1} \phi_{n-1}) \right]
\label{s10}
\end{multline}
with $\gamma=5-2(\alpha+\beta)$
defines a two-parameter family of models with 
translationally invariant kink solutions. 

The discretisation \eqref{s10} includes, as particular cases, the 
Bender-Tovbis function (\ref{nonlinearity-Tovbis}) 
(which results from setting  $\alpha = \beta = 0$)
and the Kevrekidis nonlinearity
(\ref{nonlinearity-Panos}) (for which $\alpha = \frac52$, $\beta = 0$).
Another simple function arises by letting $\alpha = \gamma = 0$; this is
a new model:
\[
Q =  \frac18 \phi_n (\phi_{n+1}^2+ \phi_n^2 + \phi_{n-1}^2 + \phi_{n+1} \phi_{n-1}).
\]

Now let $m=1$.
The most general quadratic $P_2$ satisfying (\ref{as1})--(\ref{as2}) is
\begin{equation}
P_2(u,v) = \frac12 - \alpha (u - v)^2 -
\frac12 uv, 
\label{s12}
\end{equation}
with an arbitrary $\alpha$.
Note that for $h<2$ and any $\alpha$, $\phi_*=\mp 1$ are a 
source and a sink. Hence all resulting models will
exhibit continuous families of kinks.
Substituting Eq.(\ref{s12}) for $H$ in (\ref{m3}), we obtain just a
particular case
of the nonlinearity (\ref{s10}), corresponding to the choice of the 
quartic (\ref{s8}) in the form of a complete square: $P_4=P_2^2$.
To obtain 
{\it new\/} models, we need to  note  
a symmetry $I_3(\phi_{n-1}, \phi_n, \phi_{n+1})=0$
which follows from  Eq.(\ref{s5}) with $m=1$.
Here
$$
I_3 \equiv P_2(\phi_{n+1}, \phi_n) (\phi_n-\phi_{n-1})
+P_2(\phi_n, \phi_{n-1}) (\phi_n-\phi_{n+1}).
$$
Equation (\ref{m3}) remains valid if $\beta I_3$ is added to
its right-hand side, with an arbitrary coefficient $\beta$.
The resulting function $Q$ has the form
\begin{multline}
Q = \alpha^2 (\phi_{n+1}^3+ \phi_{n-1}^3) + 2\gamma \left( \alpha
- \beta  \right)\phi_{n+1} \phi_n \phi_{n-1} \\  + \alpha \left( \alpha - \beta
\right) \phi_{n+1} \phi_{n-1} (\phi_{n+1}+ \phi_{n-1}) \\
+ \left[ 2\alpha^2 + \gamma^2 + \beta (\gamma -
\alpha) \right] \phi_n^2 (\phi_{n+1}+\phi_{n-1}) \\
 + \alpha \left(
2 \gamma + \beta\right) \phi_n(\phi_{n+1}^2 +\phi_{n-1}^2)
+
2 \alpha \left( \gamma + \beta
\right) \phi_n^3,
 \label{s14}
\end{multline}
where $\gamma = \frac{1}{2} - 2\alpha$.
Eq.\eqref{s14}   defines a two-parameter family of
 discretisations supporting
translationally-invariant kinks. 
These models cannot be obtained within 
the $m=2$ analysis above --- unless $\beta=0$, of course.

 Letting $\alpha = \beta =\frac{1}{6}$, we recover the model of  
Speight and Ward, Eq.(\ref{nonlinearity-Speight}).
Another particularly simple, new, model is obtained by taking
$\alpha=0$ and $\beta = -\frac12$:
$$
Q = \frac12 \phi_{n-1} \phi_n \phi_{n+1}.
$$

It is instructive to compare  discretisations furnished by our
 method with those arising from the requirement of the
 absence of resonant radiation from the kink \cite{BT97}.
The advance-delay equation associated with Eq.\eqref{stat},
\begin{multline}
 \phi(x+h)  - 2 \phi(x) + \phi(x-h)  \\
+h^2 f(\phi(x-h), \phi(x), \phi(x+h))=0,
\label{advance-delay}
\end{multline}
 can be solved 
to all orders as a perturbation expansion in $h$
\cite{ODI}; the resultant 
solution depends continuously on the position parameter $x^{(0)}$.  
It is, therefore, only the terms which lie
beyond all orders in $h$ which present an obstacle to the existence of a
translationally invariant kink. These terms
vanish if Stokes constants at all orders vanish.
The leading Stokes constant vanishes 
 if there 
exists a convergent solution in  powers of $z^{-1}$ to the equation
\begin{multline}
\label{innereqn}
\varphi(z+1)-2\varphi(z)+\varphi(z-1) \\ - Q(\varphi(z-1), \varphi(z),
\varphi(z+1))\big|_{h=0} = 0
\end{multline}
(see e.g. \cite{Tovbis1}).
This equation comes from a rescaling of 
Eq.\eqref{advance-delay} near the
singularities of its leading-order solution (\ref{travelling}) at 
$x_n=\pi i(1+2n)$, 
$n \in
\mathbb{Z}$, in the limit $h\to 0$.  The  convergence of a
power-series solution to Eq.(\ref{innereqn}) 
is necessary for the absence of 
oscillatory  radiation tails in its ``parent" equation
(\ref{advance-delay}).

 In general, a numerical procedure is required to
 determine whether the series converges for a given $Q$, but 
we can 
 easily
generate a class of models for which  
it 
truncates after the first term.  
This was the method employed in Ref.\cite{BT97} in deriving
Eq.\eqref{nonlinearity-Tovbis}.
It is a matter of direct substitution to check that the most
general cubic polynomial for which 
$\varphi = 2/z$ is a solution of Eq.\eqref{innereqn}, is
\begin{multline}
\label{sss} 
Q = \sigma \phi_n(\phi_{n+1}+\phi_{n-1})^2-2 \sigma \phi_{n+1}
\phi_{n-1} (\phi_{n+1}+\phi_{n-1}) \\
+\left(\frac14 -\frac{\beta}{2} \right)\phi_n^2(\phi_{n+1}+\phi_{n-1})
+\beta \phi_{n-1} \phi_n \phi_{n+1},
\end{multline}
with $\sigma, \beta$  arbitrary constants.

\begin{figure}[tbp]
\begingroup%
  \makeatletter%
  \newcommand{\GNUPLOTspecial}{%
    \@sanitize\catcode`\%=14\relax\special}%
  \setlength{\unitlength}{0.1bp}%
\begin{picture}(2339,1620)(0,0)%
\special{psfile=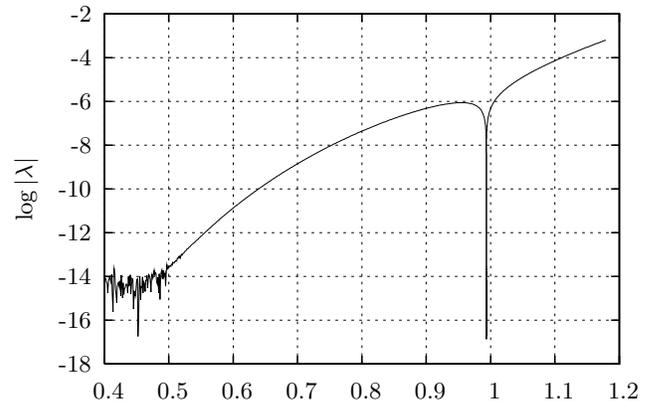 llx=0 lly=0 urx=234 ury=162 rwi=2340}
\put(50,860){%
\special{ps: gsave currentpoint currentpoint translate
270 rotate neg exch neg exch translate}%
\makebox(0,0)[b]{\shortstack{$\log|\lambda|$}}%
\special{ps: currentpoint grestore moveto}%
}%
\put(2240,100){\makebox(0,0){ 1.2}}%
\put(1997,100){\makebox(0,0){ 1.1}}%
\put(1755,100){\makebox(0,0){ 1}}%
\put(1512,100){\makebox(0,0){ 0.9}}%
\put(1270,100){\makebox(0,0){ 0.8}}%
\put(1027,100){\makebox(0,0){ 0.7}}%
\put(785,100){\makebox(0,0){ 0.6}}%
\put(542,100){\makebox(0,0){ 0.5}}%
\put(300,100){\makebox(0,0){ 0.4}}%
\put(250,1520){\makebox(0,0)[r]{-2}}%
\put(250,1355){\makebox(0,0)[r]{-4}}%
\put(250,1190){\makebox(0,0)[r]{-6}}%
\put(250,1025){\makebox(0,0)[r]{-8}}%
\put(250,860){\makebox(0,0)[r]{-10}}%
\put(250,695){\makebox(0,0)[r]{-12}}%
\put(250,530){\makebox(0,0)[r]{-14}}%
\put(250,365){\makebox(0,0)[r]{-16}}%
\put(250,200){\makebox(0,0)[r]{-18}}%
\end{picture}%
\endgroup
 
\caption{The smallest-modulus eigenvalue as a function of $h$.
The cusp occurs at the point $h=0.993$  where $\lambda$ changes sign.}
\label{eigval}
\end{figure}

The fact that the leading-order Stokes constant is zero is 
necessary but not sufficient for 
 equation (\ref{advance-delay}) to have continuous families of kinks
for finite $h$. We tested, numerically, a particular representative 
from the class (\ref{sss}):  
\begin{equation}
Q = \phi_{n+1}\phi_{n-1}
\frac{\phi_{n+1}+\phi_{n-1}}{2} - 
\phi_n \frac{\phi_{n+1}^2+\phi_{n-1}^2}{4}.
\label{olimodel}
\end{equation}
This model is obtained by letting $\sigma = -\frac{1}{4}$ and $\beta = \frac12$.
To check whether translationally-invariant 
kinks exist or not, 
 we have computed a stationary 
on-site kink for the model (\ref{olimodel}) and calculated 
eigenvalues of the associated linearised operator
for an equidistant sequence of $h$-values, ranging from $h=0$ to
$h=1.179$ with increment $0.001$. For $h$ smaller
than $0.556$, the smallest-modulus eigenvalue
was found to be smaller
than $10^{-12}$ which is 
our numerical error of computation.
 However as $h$ increases from $0.556$,
the smallest eigenvalue grows  
to $\lambda=9 \times 10^{-7}$ at $h=0.955$,
then   decreases, crosses through zero
at $h=0.993$,  after which grows in modulus to $\lambda=-6 \times 10^{-4}$
at $h=1.179$ (Fig.\ref{eigval}).  
Thus, the zero eigenvalue, indicating the existence 
of a continuous family of solutions, is not present in
the spectrum for $h>0.556$.
For $h <0.556$, 
the smallest $\lambda$ is apparently also nonzero (though very small).
The only exception is the value $h=0.993$ for which
the zero mode does exist.
This means that the model (\ref{olimodel}) supports a continuous family
of
kinks for just  one, isolated, value of $h$.

However, there exists a family of
exceptional discretisations which reduces to (\ref{sss})
in the limit $h \to 0$.
Indeed, 
when $\alpha=0$ in Eq.(\ref{s12}), the map (\ref{s5})
has one more symmetry:
$I_2(\phi_{n-1},\phi_n,\phi_{n+1})=0$, where 
\[
I_2 \equiv 
 \left( 1+\frac{h^2}{4} \right)
\phi_n (\phi_{n+1}+\phi_{n-1})
-2 \phi_{n-1} \phi_{n+1}-\frac{h^2}{2}.
\]  
We can add $\sigma (\phi_{n+1}+\phi_{n-1}) I_2$
to the right-hand side of (\ref{m3}), along with $\beta I_3$ 
(with $\sigma,\beta$ arbitrary constants.) This gives rise
to the following family of discretisations:
\begin{multline}
Q= \sigma \phi_n(\phi_{n+1}+\phi_{n-1})^2 - 2 \sigma \phi_{n-1}
\phi_{n+1} (\phi_{n+1} +\phi_{n-1}) \\
+\left( \frac14 -\frac{\beta}{2} \right) \phi_n^2
(\phi_{n+1}+\phi_{n-1}) +
\beta \phi_{n-1} \phi_n \phi_{n+1} \\
+ \frac14 \sigma h^2 \phi_n (\phi_{n+1} + \phi_{n-1})^2.
\label{s20}
\end{multline}
Except for the last, ${\cal O}(h^2)$, 
term, this coincides with Eq.(\ref{sss}).

For the map (\ref{s5})+(\ref{s12}) with $\alpha=0$, the kink
can be found explicitly. This implies that the discretisations
(\ref{s20}) also
share an explicit kink solution (for all $\beta$ and $\sigma$):
$\phi_n=\tanh (an-x^{(0)})$, with $\tanh a=h/2$.
 
Our final remark is on the 
conserved quantities of Eq.(\ref{phi-4}). The translation 
invariance of the stationary kink does not imply the invariance of 
equation (\ref{phi-4}) and hence the  conservation
of momentum. The discretisation
(\ref{m3}) [and hence (\ref{s10})] conserves  momentum
\cite{Panos} whereas the nonlinearities
(\ref{s14})  and (\ref{s20}) (with $\beta, \sigma \neq 0$)   ---  do
not. Moreover, that the discretisation $f$ is exceptional does not
guarantee that Eq.(\ref{phi-4}) has any integral of motion whatsoever.
In particular, out of the three families (\ref{s10}), (\ref{s14}),
and
(\ref{s20}), only one model conserves energy, namely Speight and Ward's,
Eq.(\ref{nonlinearity-Speight}).

In conclusion, we have identified three families of discretisations
of the $\phi^4$ equation which support translationally invariant
stationary kinks: Eqs.(\ref{s10}), (\ref{s14}) and (\ref{s20}).
In each case we have exhibited, explicitly,
  the underlying 1D map.

\acknowledgments
I.B. is a Harry Oppenheimer Fellow.
O.O. was supported by 
a Nelli Brown Spilhaus scholarship.

\end{document}